\documentclass[fleqn,usenatbib]{mnras}
\usepackage{graphics,epsf}
\usepackage{amsmath}                
\usepackage{amsfonts}               
\usepackage{amssymb}                
\usepackage{epsfig}                 
\usepackage{graphicx}
\usepackage{float}

\newcommand{\km}{{~\rm km}}
\newcommand{\s}{{~\rm s}}

\title[The final phase of common envelope]{Energizing the last phase of common envelope removal}

\author[N. Soker]{
Noam Soker$^{1}$\thanks{E-mail: \href{mailto:soker@physics.technion.ac.il}{soker@physics.technion.ac.il}}
\\
$^{1}$Department of Physics, Technion, Haifa 3200003, Israel\\
\\
}

\begin{document}
\label{firstpage}
\pagerange{\pageref{firstpage}--\pageref{lastpage}}
\maketitle

\begin{abstract}
We propose a scenario where a companion that is about to exit a common envelope evolution (CEE) with a giant star accretes mass from the remaining envelope outside its deep orbit and launches jets that facilitate the removal of the remaining envelope. The jets that the accretion disk launches collide with the envelope and form hot bubbles that energize the envelope. Due to gravitational interaction with the envelope, that might reside in a circumbinary disk, the companion migrates further in, but the inner boundary of the circumbinary disk continues to feed the accretion disk. While near the equatorial plane mass leaves the system at a very low velocity, along the polar directions velocities are very high. When the primary is an asymptotic giant branch star, this type of flow forms a bipolar nebula with very narrow waists. We compare this envelope removal process with four others last-phase common envelope removal processes. We also note that the accreted gas from the envelope outside the orbit in the last-phase of the CEE might carry with it angular momentum that is antialigned to the orbital angular momentum. We discuss the implications to the possibly antialigned spins of the merging black hole event GW170104.
\end{abstract}


\section{INTRODUCTION}
\label{sec:intro}

Numerical simulations of the common envelope evolution (CEE) in recent years that employ only the gravitational energy of the in-spiraling binary system did not achieve consistent and persistent ejection of the common envelope during the in-spiraling process (e.g., \citealt{TaamRicker2010, DeMarcoetal2011, Passyetal2012, RickerTaam2012, Nandezetal2014, Ohlmannetal2016, Staffetal2016MN8, NandezIvanova2016, Kuruwitaetal2016, IvanovaNandez2016, Iaconietal2017b, DeMarcoIzzard2017, Galavizetal2017, Iaconietal2017a}, and many more references to earlier papers within these papers).

In light of these difficulties theoretical studies have been searching for an extra energy source in addition to that of the gravitational energy released by the in-spiraling binary system, {{{{ e.g., \cite{Kruckowetal2016} for a recent list of some extra energy sources. }}}}  One such extra energy source is the recombination energy of the ejected envelope (e.g., \citealt{Nandezetal2015, IvanovaNandez2016, Kruckowetal2016} for recent papers).
\cite{IvanovaNandez2016} identify four types of ejection processes. They term these the pre-plunge-in ejection, the outflow during the plunge-in, the outflow driven by recombination, and the ejection triggered by a contraction of the circum-binary envelope. They find that all processes might significantly contribute to the mass loss process.
They took recombination energy into consideration in their simulations, but ignored the arguments raised by \cite{SokerHarpaz2003}. In particular, when recombination takes place the optical depth gets low, and radiation efficiently escapes \citep{Harpaz1998}. In a recent study \cite{Sabachetal2017} support the conclusion of these two studies that radiation remove most of the recombination energy.

The other energy source that is considered in helping the ejection of the envelope is the accretion energy that is channeled to jets.
\cite{ArmitageLivio2000} and \cite{Chevalier2012} studied CE ejection by jets launched from a secondary star that is a neutron star, but they did not consider jets to be a general common envelope ejection process.
We adopt the view that in many cases, in particular for main sequence secondary stars spiraling-in inside the envelope of a giant star, jets facilitate the removal of the common envelope
(e.g., review by \citealt{Soker2016Rev} and a recent paper by \citealt{MorenoMendezetal2017}). Even if a fully developed accretion disk is not formed around the main sequence star (see for example \citealt{Murguiaetal2017}), but rather an accretion belt is formed, the high accretion rate through the accretion belt might lead to the launching of jets \citep{Shiberetal2016, SchreierSoker2016}. The claim made by \cite{BlackmanLucchini2014}, based on the momenta of bipolar pre-PNe, that strongly interacting binary systems, most likely in a CEE, can launch energetic jets is compatible with the suggestion that jets can help remove the common envelope.

As jets remove material from their surrounding they reduce the accretion rate and hence also reduce their own intensity. This is a negative feedback mechanism. There is also a positive feedback effect. The jets remove angular momentum, high-entropy gas, and energy from the vicinity of the accreting secondary star, and by that reduce the pressure and allow the accretion \citep{Shiberetal2016, Staffetal2016MN}, most likely through an accretion disk or an accretion belt. If not for this positive feedback effect, accretion would be much lower due to the build up of a high pressure in the vicinity of the accreting secondary star (e.g. \citealt{RickerTaam2012, MacLeodRamirezRuiz2015, Murguiaetal2017}).

When the jets efficiently remove the envelope from the beginning of the in-spiraling phase, the system does not enter a CEE, at least not in the initial phase. The system rather experiences the recently suggested process of grazing envelope evolution (GEE; \citealt{SabachSoker2015, Soker2015, Shiberetal2017, ShiberSoker2017}). In the GEE the secondary star that orbits near the very outer regions of the envelope accretes mass from the envelope and launches jets, such that these jets manage to eject the envelope gas along the orbit of the secondary star.
For high mass accretion rates the jets can carry energy that surpass the recombination energy of the ejected gas, and when the primary star is a giant star, the kinetic energy of the jets can be  larger than the gravitational energy that is released by the in-spiraling secondary star. In extreme cases the secondary star will stay at a very large orbital separation while removing the envelope with the jets it launches \citep{Soker2017}.

But even with these extra energy sources, it is likely that in the last phase of the CEE the secondary star slowly spirals-in over a relatively long time. The removal of the remaining envelope occurs over a relatively long time, longer than the dynamical time of the AGB star. This is the last phase of common envelope removal. Four processes are discussed in the literature to remove the common envelope in this last phase.

In the first last-phase process the secondary star excites p-waves in the giant envelope \citep{Soker1992, Soker1993}. These waves turn to large perturbations on the surface of the giant star and enhance the mass loss rate, in particular near the equatorial plane. Even brown dwarfs and massive planets can substantially enhance mass loss rate.

According to the second last-phase process the relatively rapidly rotating envelope loses mass at a high rate, due to enhanced dust formation, while the secondary star, even a brown dwarf, orbits deep inside the envelope \citep{Soker2004}.

The third last-phase process is based on the formation of a thick circumbinary disk around the binary system (the core and the secondary star) at the final phase of the CEE (e.g., \citealt{Kuruwitaetal2016}). \cite{Kashisoker2011MN} proposed that the gravitational interaction of the binary system with this circumbinary disk transfers energy to the disk, and as a result of that the disk losses mass at a high rate.

According to the fourth last-phase process the energy released by frictional dissipation of the secondary orbit inflates the envelope and leads to  large-amplitude pulsations that enhance mass loss rate \citep{Claytonetal2017}.

In the present study we develop a fifth process to energize the final removal of the common envelope. It is based on jets launched by the secondary star as it accretes mass from the circumbinary envelope (section \ref{sec:flow}). A preliminary setting of this process can be found in \cite{Soker2014b} and in \cite{Soker2016Rev}. In section \ref{sec:last} we compare the five last-phase processes. We summarize in section \ref{sec:summary}.

\section{JETS AT THE FINAL CEE PHASE}
\label{sec:flow}
\subsection{A circumbinary disk}
\label{subsec:disk}

After the secondary star enters the common envelope it suffers a fast spiraling-in phase that is termed the plunge-in phase (see review by \citealt{Ivanovaetal2013}). At the end of this phase most of the remaining common envelope resides outside the orbit of the core and the secondary star around their mutual center of mass. The envelope has now a highly oblate structure, or even a torus like structure that we can term a circumbinary disk, as \cite{Kuruwitaetal2016} demonstrated by their hydrodynamical numerical simulation. \cite{Kashisoker2011MN} suggested that gravitational (resonant) interaction between the binary system (the core and secondary star) and the circumbinary disk causes further spiraling-in of the binary system, in a process that resembles to some degree the migration of a planet in a proto-planetary disk. \cite{ChenPodsiadlowski2017} adopted the idea of \citealt{Kashisoker2011MN} and further developed it.

The gravitational (resonant) interaction between the binary system and the circumbinary disk transfers energy and angular momentum from the in-spiraling binary system to the disk. The disk expands and radiates at a very high rate. According to the calculation of \cite{Kashisoker2011MN} the in-spiraling binary system powers the disk with about 10 times the Eddington luminosity. The disk cannot radiate it, and must expand and blows an intensive wind. \cite{ChenPodsiadlowski2017} find a slower decay rate. Also, some extra energy can be carried out by the efficient convection in the circumbinary disk and then be radiated away.
In any case, the gravitational interaction of the binary system and the circumbinary disk (or circumbinary envelope) increases the mass loss rate in the last phase of the CEE.

Some bound mass does not reside in the circumbinary disk, but rather falls from large distances back toward the center \citep{Kuruwitaetal2016}. Fraction of this mass fills the entire volume, including along the polar directions (along he orbital angular momentum axis).

The gas in the envelope that now resides in a circumbinary disk is not fully supported by the centrifugal force, and mass flows toward the center. The secondary star accretes mass from this inflow through an accretion disk. Not only the accretion process releases gravitational energy in radiation and jets, but it also removes mass from the very inner regions of the remaining envelope. This mass has the highest binding energy per unit mass in the remaining envelope. So the accretion of mass onto the secondary star facilitates envelope removal in two ways.

Consider an element of gas of mass $\Delta m_{\rm acc}$ in the circumbinary disk that resides at radius $a_i$ from the center of mass, and due to the gravitational interaction and friction in the disk it is accreted onto the secondary star at an orbital radius $a_2 < a_i$ from the center of mass.
It releases an accretion energy (defined positively) of
\begin{equation}
\begin{split}
\Delta E_{\rm acc} & \simeq  \frac{1}{2} \frac {G M_2 \Delta m_{\rm acc}} {R_2} \\
&+ \frac{1}{2} G (M_{1c} + M_2) \Delta m_{\rm acc} \left( \frac{1}{a_2}-\frac{1}{a_i} \right),
\end{split}
\label{eq:de}
\end{equation}
where $M_{1c}$ here is the mass of the core of the primary star.
When the mass is accreted to the secondary star it is also removed from the envelope, and hence reduces the binding energy of the envelope by an amount of $\Delta E_{\rm bind} \simeq (1/2) G(M_{1c}+M_2) \Delta m_{\rm acc} /a_i$.  When counting the contribution of the accretion of the mass  $\Delta m_{\rm acc}$ to envelope removal at the last CEE phase, we should add $\Delta E_{\rm bind}$ to the total energy released by this mass as it is accreted from radius $a_i$ to radius $a_2$ and on to the secondary star.
The energy contribution to envelope removal of the accretion of mass $\Delta m_{\rm acc}$ in the the last-phase of common envelope removal is then
\begin{equation}
\begin{split}
\Delta E_{\rm cont} &= \Delta E_{\rm acc} + \Delta E_{\rm bind}  \\
&\simeq \frac{1}{2} \frac {G M_2 \Delta m_{\rm acc}} {R_2} +
\frac{1}{2}  \frac {G (M_{1c} + M_2) \Delta m_{\rm acc}}{a_2}.
\end{split}
\label{eq:deacc}
\end{equation}
The contribution of the two last terms in equation (\ref{eq:deacc}) to envelope removal at the last common envelope phase might be of the same order of magnitude as far as energy is concerned. But the accretion on to the secondary star might result in launching jets, and this we consider to be the most significance effect of the accreted mass.

\subsection{Jets}
\label{subsec:jets}

A preliminary suggestion that the secondary that accretes mass from the circum-binary disk launches jets that facilitate the common envelope removal can be found in the paper \cite{Soker2014b}.
The accretion disk around the secondary star is expected to launch jets. The velocity of the jets is about the escape velocity from the main sequence star, $v_j \simeq 500 \km \s^{-1}$. The angle of the symmetry axis of each jet relative to the orbital angular momentum axis is $\tan \theta_t = v_j/v_{\rm orb}$ where $v_{\rm orb}$ is the velocity of the secondary star around the center of mass of the binary system. Scaling for $M_{1c}=1M_\odot$ and $M_2=0.3 M_\odot$ and for a circular orbit with an orbital separation of $a$ we derive
\begin{eqnarray}
\tan \theta_t = \frac {v_{\rm orb}}{v_j} =
0.34
\left( \frac {v_j}{500 \km \s^{-1}} \right)^{-1}
\left( \frac {M_{1c}}{1 M_\odot} \right) \nonumber \\
\times
\left( \frac {M_1+M_2}{1.3 M_\odot} \right)^{-1/2}
\left( \frac {a}{5 R_\odot} \right)^{-1/2} .
  \label{eq:thetat}
\end{eqnarray}
This value corresponds to $\theta_t = 19^\circ$.

The orbital motion tilts the two jets in the same direction, both jets are tilted parallel to the orbital motion of the secondary star.
To the tilt angle we need to add the initial width of the jets at launching. Over all, the leading part of the jets (toward the direction of orbit) can be tilted to the axis of the orbital angular momentum by tens of degrees. As the circumbinary disk is thick, it is very likely that the leading parts of the jets, toward the orbital motion, interact with the inner boundary of the circumbinary disk.

{{{{ The important factor for the jets is that they deposit their energy in the ambient medium. For that, the jets must not drill a hole and escape out from the system (e.g., \citealt{Soker2016Rev}). This condition is fulfilled if the jets are sufficiently wide, and/or the jets precesses rapidly enough, and/or the source of the jets moves sufficiently fast relative to the ambient gas. In the case of the CEE the orbital motion of the secondary star, that is the source of the jets, is sufficient to prevent penetration, in particular for main sequence secondary stars (e.g., \citealt{Papishetal2015}). So for the scenarios discussed in the present study, the jets need not be wide to inflate large hot bubbles.  }}}}

The jets can also interact with the inflowing gas. As gas is expected to fall from all directions, the parts of the jets that do not interact with the circumbinary disk will interact with inflowing gas. As the jet gas is shocked, whether interacting with the gas along polar directions or denser gas in the circumbinary disk, the shocked gas inflates hot bubbles. The hot bubbles buoyantly rise in the radial direction and energize the circumbinary envelope \citep{Sabachetal2017}.

\subsection{Antialigned spins}
\label{subsec:spin}

During most of the CEE the mass residing inner to the secondary star is denser than the mass laying outside the secondary orbit. The secondary star accretes mass with a positive angular momentum. Namely, the angular momentum of the accreted mass with respect to the center of mass of the secondary star is aligned with the orbital angular momentum.
When the secondary star reaches very small orbital separation and a circumbinary disk is being built, the envelope density outside the orbit of the secondary star might be larger than the density of the gas inner to the secondary orbit. At the initial stage the toroidal circumstellar envelope (or circumbinary disk) has less than the Keplerian angular velocity. Namely, it is not supported by centrifugal forces and its orbital velocity and angular momentum around the primary star is less than that of the secondary star. This situation is not stable, as mentioned in section \ref{subsec:disk}, and mass from the toroidal circumbinary envelope flows inward. At this stage the secondary star accretes mass with a negative angular momentum, i.e., the angular momentum is antialigned with the orbital angular momentum.

The general conditions for the accretion of mass with antialigned (to the orbital angular momentum) angular momentum (relative to the accreting star) are
\begin{equation}
\left( \frac{d \rho_{\rm e}}{dr} \right)_{r=a_2} >0,
\quad {\rm and} \quad
j_{\rm e} (a) < j_2.
  \label{eq:antia}
\end{equation}
The first condition states that the envelope density, $\rho_{\rm e}$, outside the orbit should be larger than that inner to the orbit, and the second condition is that the specific angular momentum of the accreted gas, $j_{\rm e}$, around the center of mass of the binary system is less than that of the accreting star around the center of mass of the binary system, $j_2$.
When the system reaches the last phase of the CEE, these conditions might be fulfilled. The accretion will spin-down the star. However, it is not clear that the amount of accreted antialigned angular momentum at the end of the CEE is sufficient to flip the spin, such that the star spin will be antialigned to orbital angular momentum.

The situation might be different if the primary star is a massive star that undergoes a core collapse supernova (CCSN) explosion that forms a massive black hole. In that case the collapse of the rapidly rotating core forms a central object, first a neutron star and then a black hole, that is expected to launch well collimated jets, that expel mass only along the two polar directions \citep{GilkisSoker2016}. \cite{GilkisSoker2016} also argue that the removal of core and envelope mass along the polar directions reduces the gravity, and equatorial gas expands and part of it falls back. This fall back gas has a lower than Keplerian angular momentum. If the removal of little mass does not destroy the binary system, the fall back gas that is accreted by the secondary star and/or the central black hole might have an antialigned angular momentum. If the secondary star is a black hole itself, the black hole binary system might end with a spin angular momentum that is antialigned to the orbital angular momentum.

These proposed two speculative processes, last-phase of the CEE and post-CCSN fall back, might imply that the new observations of close to zero spins, or even a possible antialigned spins, in the merging black hole event GW170104 \citep{Abbottetal2017} might not rule out formation of binary black holes via a CEE (for such scenarios see, e.g., \citealt{Belczynskietal2016}). {{{{ This preliminary speculative claim must be studied in more details. }}}}

\section{THE LAST-PHASE PROCESSES}
\label{sec:last}

In this section we summarize the five processes that we aware of that have been proposed to facilitate the final envelope removal during the last phase of the CEE. In this last phase the secondary star has given up most of its initial orbital angular momentum, but it yet to release more orbital gravitational energy than it released till the beginning of the last phase. For example, in the last phase it spirals-in from $a \approx 10-20 R_\odot$ to $a_f \simeq 2-5 R_\odot$.  The last phase lasts for a time longer than the dynamical time of the giant star.

In Table \ref{Tab:Table1} we list the five last-phase common envelope removal processes, the last one being the one proposed here in section \ref{subsec:jets}. It is important to emphasize that these processes do not exclude each other. It is even expected that some of them operate simultaneously. {{{{ At the termination of the CEE the star is expected to be engulfed in a dusty wind. It will be hard to directly observe the star. Consequences of the different processes might be observed at later times, as we discuss here. }}}}
\begin{table*}
\begin{center}
\begin{tabular}{|l|l|l|l|l|}
  \hline
Process    & Energy  & Mass loss & Comments & Reference \\
\hline
\hline
p-waves    & Binary  & Surface        &$\bullet$ Also works for & \cite{Soker1992} \\
           &+ nuclear& perturbations  &$~$ BDs and planets     &                 \\
 \hline
Rotation   & Nuclear & $\vec B$ + Enhanced&$\bullet$ With other processes&\cite{Soker2004} \\
           &         & dust formation     &$\bullet$ A slow process  &\cite{Soker2004IAUS}                \\
 \hline
Circumbinary& Binary & Binary-disk    &$\bullet$ Efficient   & \cite{Kashisoker2011MN}\\
 disk      &         & interaction    &$\bullet$ More spiral-in &  \\
\hline
Envelope  &  Binary      & Pulsation +  & $\bullet$ Similar to  & \cite{Claytonetal2017}  \\
Inflation &+recombination& recombination& $~$ pre-CCSNe outbursts   &  \\
\hline
 Jets     &Accretion & Jet-envelope   &$\bullet$ Secondary spin might& \cite{Soker2014b} \\
          &          & interaction    & $~$ end antialigned & and this paper\\
\hline
\end{tabular}
\end{center}
 \caption{Comparing five processes that have been suggested to enhanced mass loss rate at the last phase of the CEE. The energy sources and processes are as follows. Nuclear: the thermal energy of the envelope and the luminosity as supplies by the nuclear burning in the core. $\vec B$: magnetic fields in the envelope that are amplified by a dynamo. Binary: The gravitational energy that is released by the in-spiraling binary system. Abbreviation: BDs: brawn dwarfs. CCSNe: core-collapse supernovae.    }
 \label{Tab:Table1}
\end{table*}

\emph{p-waves.} In this process the secondary star orbits deep inside the envelope. It excites p-waves that propagate to the lower density regions in the outer envelope, where the relative amplitude increases \citep{Soker1992, Soker1993}. The relative pressure amplitude in the equatorial plane near the surface of the giant star, at radius $R_1$, is given by the expression (eq. 5.7 in \citealt{Soker1992})
\begin{eqnarray}
\left[ \frac {\vert P^\prime \vert }{P} \right]_{r\simeq R_1} \approx
5 \frac {M_2}{0.2 M_1(a)}
\left( \frac {a}{0.1 R_1} \right)^{-0.75},
  \label{eq:pressurep}
\end{eqnarray}
where $a$ is the orbital separation and $M_1(a)$ is the mass of the primary star inner to the orbit of the secondary star.
Of course, the highly nonlinear perturbation does not obey the assumptions that lead to its derivation. The expression was derived for brown dwarfs and planets. But here we substitute values for a stellar companion to emphasize the non-linear regime that these waves reach. In the nonlinear regime they dissipate their energy in the envelope and lead to its expansion \citep{McleySoker2014}. Both the expansion and high pressure perturbations on the surface of the giant star increase the mass loss rate.

The energy comes from the orbital energy, since as the secondary star excites waves it spirals-in. When the secondary object is a brown dwarf or a massive planet, then perturbations are smaller, and the radiation of the giant star (that comes from nuclear burning) becomes an important source of energy to eject mass from the surface of the giant, where dust formation in large quantities is expected.

{{{{  The excitation of p-waves is expected to lead to very clumpy mass loss geometry. Although the mass loss process of single AGB stars is clumpy, the p-waves might form massive-large clumps of dusty wind. This process requires further calculations before comparison to nebulae around post-CEE binaries can be done. }}}}

\emph{Rotation.} The more rapidly rotating giant star, due to the initial orbital angular momentum of the binary system, amplifies magnetic fields (e.g., \citealt{Nordhausetal2007}) that might lead to surface activity that leads to enhanced dust formation (e.g., \citealt{SokerClayton1999}), hence high mass loss rate \citep{Soker2004}.  This process simply uses the luminosity of the giant star, i.e., the nuclear energy in the core, as in the single star mass loss process, but with enhanced dust formation. It is a slow process, but might work alongside all the other four proposed processes. In general, under the process of rotation we also include any amplification of magnetic fields in the AGB envelope (e.g., \citealt{Soker2000, Nordhausetal2007, Ohlmannetal2016b}).
{{{{ There is no late imprint that is unique to this process. Strong local magnetic fields in the wind might hint that this process took place. As well, if a planet was engulfed and the mass loss rate is small such that the dust does not obscure the star, then cool or hot spots might also hint on magnetic activity due to a dynamo that operates in a rotating envelope.   }}}}

\emph{Circumbinary disk.} Gravitational (resonant) interaction transfers orbital energy to the envelope that resides outside the orbit, possibly in a circumbinary disk \citep{Kashisoker2011MN}. Part of this energy is dissipated locally, and then carried out as radiation or by convection, and part of it initiates disk-outflow \citep{Kashisoker2011MN}. If there is only a circumbinary disk, as expected in the very final stage of the last CEE phase, then this is the only process to transfer energy from the binary outward. In most of the time, however, there is gas around the secondary star, and there are also the channels of p-waves excitation and frictional dissipation.
{{{{ The effect of a circumbinary disk might be a slowly expanding dense equatorial outflow with an age smaller than that of the main nebula that was ejected during the CEE. This will be  most likely accompanied by jets (as the secondary star accreted from the circumbinary disk). Jets younger than the main nebula might be a strong signature of final envelope removal by jets (the fifth process listed in Table \ref{Tab:Table1}.)   }}}}

\emph{Envelope Inflation.} \cite{Claytonetal2017} study the effect of energy deposition at the base of the convective envelope of a giant star. They attribute the energy to frictional dissipation (local gravitational interaction), but as mentioned above, resonant interaction and p-waves can also dissipate energy to the envelope. \cite{Claytonetal2017} find that the energy deposition in the deep envelope leads to the expansion of the envelope and to large-amplitude pulsations that enhance mass loss rate.

\cite{McleySoker2014} study the response of massive stars to energy deposition in their envelope months before they explode as core collapse supernovae. The expansion process of the giant in the simulations of
\cite{Claytonetal2017}, e.g., their figure 8, is similar in some respects to the expansion of the envelope of massive stars before they explode (figure 5 of \citealt{McleySoker2014}).
{{{{  A signature of intermittent envelope inflation might be strongly variable mass loss rate, with several dense shells in the nebula that were formed by high mass loss rate during envelope inflation phases.}}}}

We also note that for the parameters used by \cite{Claytonetal2017}, i.e., a secondary star of mass $0.3 M_\odot$ orbiting at the base of the convective envelope of a giant of radius of $100 R_\odot$, equation (5.7) of \cite{Soker1992} as presented here in equation (\ref{eq:pressurep}), predicts pressure disturbances on the surface that are tens of times the unperturbed pressure. Namely, the effect of p-wave excitation cannot be neglected with respect to frictional dissipation during the slow spiraling-in phase. In many cases the two processes are expected to operate simultaneously.

We summarize this section by emphasizing that before we better understand the envelope removal in the last phase of the CEE, one must consider all the processes that are listed in Table \ref{Tab:Table1}. One should not study in isolation only one of these processes and then claim for solving the CEE.

\section{SUMMARY}
\label{sec:summary}

We developed the idea \citep{Soker2014b} that jets might contribute to the removal of the remaining common envelope in the last phase of the CEE (section \ref{subsec:jets}). The jets are launched by the secondary star that accretes mass from the circumbinary envelope (section \ref{subsec:disk}). We list this process with the four other last-phase common envelope removal processes that have been proposed in the literature (Table \ref{Tab:Table1}).
Before the community can follow the entire CEE, from start to end, and determine the most important processes that remove the common envelope, it is imperative to consider all the processes that are listed in Table \ref{Tab:Table1}.

Adding our proposal that jets facilitate the removal of the envelope in the last phase of the CEE to the grazing envelope evolution (GEE) phase that takes place before the CEE, we raise the  possibility that a large fraction of the envelope is removed by jets at the very beginning and at the end of the CEE. During the CEE phase where most of the spiraling-in takes place (plunge-in) only a fraction of the envelope is removed from the system. The orbital energy deposited during the plunge-in phase is very large and plays a major role in inflating the envelope and overcoming the binding energy of the envelope, but it does not remove the entire envelope by itself.

There is one interesting outcome of mass accretion from a sub-Keplerian circumbinary envelope. When the accreted gas has a specific angular momentum around the center of mass of the system that is less than the specific orbital angular momentum of the accreting object, then the direction of the angular momentum around the center of the accreting object is opposite to that of the orbital angular momentum (eq. \ref{eq:antia}). The accretion of antialigned angular momentum might flip the spin of the accreting objects. This effect might be large when the primary star is a massive star that explodes as a supernova and forms a black hole (section \ref{subsec:spin}). This implies that the formation of a binary black hole system might lead to antialigned spins of one or two objects, or spins close to zero.

I thank Amit Kashi for helpful comments, {{{{ and an anonymous referee for comments that improved the presentations of the results. }}}} This research was supported by the Israel Science Foundation.

\label{lastpage}
\end{document}